# Low Losses Left Handed Materials Using Metallic Magnetic Cylinders.

**N. García and E.V. Ponizovskaia**

Laboratorio de Física de Sistemas Pequeños y Nanotecnología,
Consejo Superior de Investigaciones Científicas ,
Serrano 114, 28006 Madrid, Spain

*Abstract*

We discuss materials based on arrays of metallic magnetic cylindrical structures near ferromagnetic resonance with applied magnetic fields at microwave frequencies. We have found that the materials have a negative refraction index when the appropriate structure is chosen. Numerical FDTD simulations were performed, after a very large number of geometries were swept. The simulations reveal that only ferromagnetic cylinders, with diameters of 0.1 cm and 0.5 cm apart, and with periodic or random configurations, are left-handed materials with very small losses; i.e. with transmitivity practically unity or no losses.

Recently, there have been reports of man-made materials called meta-materials, formed by periodic arrays of cylinders and split ring resonators [1], which may exhibit a negative refraction index [2,3]. These materials, whose properties were theoretically predicted by Veselago [4], are known as left-handed media (LHM). The materials that were reported so far, however, exhibited significant losses and a small transmission of the order or smaller than $10^{-3}$, meaning there is a significant problem with propagation in these systems [5]. Utilizing practically the same meta-materials, other experiments showed much larger transmission. In these cases, however, the claimed permittivity and permeability had negative imaginary parts [6]. Regardless, the realization of these LHM materials is certainly interesting. It is, furthermore, convenient to obtain them in other ways with very large transmissions, so that the propagation is not a problem, and that if possible the real parts of the permittivity and permeability are negative, while the imaginary parts are positive.

It is well known that at ferromagnetic resonance at right or linear polarized radiation there is a frequency region where the real part of the permeability is negative and its imaginary part is positive with a small value (see Fig. 1a) [7]. In addition, for s-polarization radiation, an array of cylinders can exhibit a region where the same thing happens with the permittivity when the diameter and the separation between cylinders (see Fig.1b), are adequately chosen. We, therefore, have the ingredients to fabricate LHM materials. The only problem is if it is possible to choose the frequencies and geometrical structures that will allow low losses and transmitivities near to one. Notice, for example, that a film of ferromagnetic metal will exhibit the real part of the refraction index as negative, while the losses are at microwaves. For a thickness, however, of the order of the wavelength of the radiation (? 1.5cm), the transmitivity will be practically zero so that there is only attenuation of the radiation.

In this paper, we take the opportunity given by the properties of magnetic materials near ferromagnetic resonance (FMR) to elaborate on the above ideas, and discuss materials based on a structure consisting of metallic magnetic cylinders under the influence of magnetic fields. A LHM has negative permittivity $\varepsilon(\omega)$, permeability $\mu(\omega)$ and refractive index $n(\omega)$, where $\omega$ is the frequency. The LHM is, moreover, a dispersive medium [4] with Re($n(\omega)$)<0 and an absorption that is determined by Im ($n(\omega)$)>0. The propagation of the electromagnetic wave in LHM exists only if the Im ($n(\omega)$) is small enough, and the damping is less then 1/e in thickness (of the order of magnitude or larger than the wavelength of the radiation).

We study the propagation in the frequency region where ferromagnetic resonance exists; i.e., f< 1-100GHz (*w=2pf*) . In the proximity of resonance $\mu$ changes sign, and becomes zero at a certain frequency $\omega_c$. For the small region of frequencies around $\omega_c$ we have Re($n(\omega)$) ≈0 and Im ($n(\omega)$) to be very small. A similar effect takes place in dielectric behavior [5]. The metallic structure has a negative permittivity in the same range of frequency. Our calculations show that there is a region of frequencies where the dielectric permittivity and the magnetic permeability are both negative and the structure has LHM properties.

We consider an array of magnetic metallic wires embedded in a dielectric medium. In our case, this is air to prevent losses due to the dielectric. We then simulate the electromagnetic wave propagation in a structure consisting of periodic or random rows of metallic cylinders. The metallic cylinders were assumed to have a frequency dependent dielectric permittivity

$$\varepsilon(\omega) = 1 - \frac{\omega_p^2}{\omega^2 + i\gamma\omega} \quad (1),$$

where $f_p = 9$ eV is the plasmon frequency, and $\gamma = 0.1$ eV is the damping frequency taken to fit the experimental values of the permittivity [8]. Notice that in the range of microwave frequencies for metals typical values are $\varepsilon(?) \sim -5000i + 10^7$, which is general for all good metals. The transmission of the considered structure should have a cut-off frequency at about $f_c = 2c/(a-d)$, where $c$ is the velocity of light, $a$ the separation between the cylinders and $d$ their diameter [5].

The permeability, however, has a resonant behavior for ferromagnetic wires at the microwave frequency range discussed above. The frequency dependent permeability at ferromagnetic resonance for right-hand radiation is:

$$\mu(\omega) = 1 - \frac{\omega_m(\omega_0 + \omega - i\alpha\omega)}{\omega^2 - (\omega_0 - i\alpha\omega)^2} \quad (2),$$

where $\omega_m$ is the effective characteristic frequency of the ferromagnet at magnetization (M), $\omega_0$ is the resonant frequency, $\gamma_M$ being the gyromagnetic ratio, $\alpha$ representing the friction coefficient [7]. The resonant frequency $?_0$ depends on the demagnetizing field [7,9]. In addition, for fields perpendicular and parallel to the cylinders, $?_0$ takes the values $?_M(H(H-2\pi M))^{1/2}$ and $?_M(H+2\pi M)$, where H is the applied magnetic field. By changing the magnitude and direction of the field we can locate the resonance in a wide range of frequencies. This is favorable because it gives a lot of versatility for the frequency in which we want to have the LHM with the accompanying negative refraction.

In Fig.1a one can see the real and imaginary parts of permeability vs. frequency $f=\omega/2p$. The imaginary part has a large peak at the frequency $\omega_0$, and the real part is negative in the region of frequencies between $\omega_0$ and the cut-off frequency (approximately around $\omega_c \approx (\omega_m + \omega_0)$). The value of the imaginary part, furthermore, depends on the friction coefficient $\alpha$, while the width of the peak at the frequency $\omega_0$ is greater then the height of $\alpha$. In Fig.1a, the permeability for $\omega_m/2p = 6GHz$, $\omega_0/2p =18GHz$ and $\alpha=0.001$ is shown. Near the cut-off frequency $\omega_c$, $Im(\mu)$ is small enough for reasonable values of the friction coefficient ($0.001<\alpha<0.01$), and the latter yield typical resonance widths of 0.09 to 0.4 GHz [7,9]. Thus, in the region where $\mu\leq0$ and $\varepsilon$ is a negative value we could have very low losses and a pronounced peak of transmission.

For our simulation we used a frequency dependent Finite-Difference Time-Domain (FDTD) method [5,10]. We model the incident wave as an s-polarized pulse. The periodic boundary conditions were used along the wave propagation direction, and the absorption boundary conditions were used in the direction perpendicular to the wave propagation. The calculations were done with grid *dx=dy=a/50=0.01cm* and time step *dt=5.6\*10$^{-4}$ a/c*, where *c* is the light velocity. Satisfactory convergence tests have been carried out calculating the propagation for different grid spacing.

Now we have to simulate our model for the best inter-cylinder distance and cylinder diameter. We choose *a=0.5cm, which gives the cut-off frequency and scale at the values we want for the LHM region.* The value d of the cylinder diameter was between 0.001 cm and 0.2 cm. This variable is the key ingredient in the modeling, and the value chosen in the frequency region of interest (18 to 20 GHz) was the one resulting in small losses and a greater degree of transmission [11]; i.e., *d= 0.1 cm.* The structure has an effective negative permittivity with a low value for the imaginary part in a wide range of frequencies Fig. 1b. To calculate the effective $\varepsilon$ we studied the metallic structure with the dielectric permittivity described by Eq.1 and permeability $\mu$ equal to 1 everywhere, i.e. out of the FMR condition. As one could see from Fig.1b, the transmission has a cut-off frequency of about 25GHz, and in the near region below, the effective dielectric permittivity is negative and its imaginary part is negligible.

We, subsequently, studied the transmission of the structure with ferromagnetic resonance where permeability inside the wires is taken as frequency dependent according to Eq.2. The values used for the parameters have been explored in a wide range and the best transmitivity was obtained when the reflectivity was practically zero because absorption is always small. This happens when the refraction index is close to -1 as is the case for $\omega_m/2p = 23$GHz and $\omega_0/2p = 18$GHz and $\alpha=0.001$. For three rows of the cylinders, the results are shown at Fig.2a for a periodic structure of the aforementioned structures for the normal incident wave. We have a big transmission in the small region around 19GHz. This is where the permittivity $\varepsilon$ and permeability $\mu$ are both negative, while the losses that are determined by the imaginary parts of $\varepsilon$ and $\mu$ are small. Thus, we have a left-handed material with small losses in this area, and a transmission of practically unity (0dB). The simulations were also done for an angle of incidence equal to $45^0$ for the periodic structure (Fig 2b), and for random cylinders with an average distance of *a=0.5 cm* with a normal incident wave (Fig.2c). The same behavior was observed in these cases as well. This is important because it shows the isotropic behavior of the material, no matter what the incident angle is the peak appears at the same point. All the results are similar and show a very high transmission near $\omega_0$ where the both $\mu$ and $\varepsilon$ are negative. This also gives versatility in the sense that we do not need a periodic structure. That is to say, hanging wires of ferromagnetic material, placed randomly with an average separation of 0.5cm will present LHM properties.¡

The increase of the friction coefficient $\alpha$ decreases the transmission a little. The region with a high transmission can be, moreover, controlled by the value of applied magnetic field *H* that determines the value of $\omega_0$. Still, the significant change of $\omega_0$ reduces the transmission. In Fig.3a the transmission for $\alpha=0.005$ is shown. One can see that it is 1.5 dB less than for $\alpha=0.001$. In order to prove that we have good propagation we have calculated the transmission of the structure while increasing the number of rows up to 10. We find that the peak narrows down, but has a transmitivity of unity for a well-determined frequency around 19GHz (Fig. 3b). The frequency $\omega_0$ determines the position of the transmission peak. We made the same calculations for $\omega_0$ equal to 15GHz and 10GHz. As a result, we had a corresponding transmission peak correspondingly near 15 or 10 GHz, though its value was a little smaller for $\omega_0$ equal to 15 and significantly smaller for 10GHz. The peak occurs because the dielectric permittivity has a greater negative value while it's imaginary part is greater as well (see Fig.1). The losses are exponentially dependent on the value $(Im(\varepsilon)Re(\mu)+Im(\mu)Re(\varepsilon))$, and these decrease as $\omega_0$ is shifted toward of the cut-off frequency, although such a structure can be created for another frequency-range by changing the cut-off frequency that depends on the inter-distance constant $a$ and the cylinder diameter $d$. In other words, we can simulate the LHM in the desired frequency region by changing the field H and the inter-distance separation between cylinders. By reducing the value of $?_m/2p$ to 6 GHz we obtain a transmitivity of 0.1 and a reflectivity of 0.9, with practically no absorption. The refraction index =-0.06 in this case [11] that is the reason for the calculated reflectivity.

One more test is that a LHM should present focusing properties as was previously discussed [4]. We have simulated the transmitivity and focusing properties of a point source at 1cm from our material, at 19GHz. Figure 4a shows the results for $?_m/2p$ =23GHz of characteristic magnetization with a clear focus at a 0.65cm distance that corresponds to a refraction index of n=-1.1, with a very small imaginary part because the transmitivity is near to unity and the reflectivity practically zero.

Another interesting aspect from our simulations, in the wide region of parameters used, is that there is a good region for LHM with wires of approximately 0.001cm in diameter. In this region, however, our calculations are not too precise due to the large ratio between the wavelength of the radiation and the wire diameter [11]. In addition, similar results can be obtained using spheres instead of wires, but this case requires considerably more difficult to construct in order to have then hanging in air to reduce losses of the medium and to simulate [11]. But can be fabricated also with the medium different of air, then allow another parameter into the manipulation to prepare the material. This may be also positive since there is large experience in composite materials.

In conclusion, we have studied the transmission properties of composite metallic media according to the behavior of their permeability and the corresponding reflection index. We have shown the existence of interesting properties at microwave frequencies in the FMR region where $Re(\mu(\omega)) <0$, and where the media has real values of *n* negative, in our case -1.1 as obtained from focusing properties. This allows us to obtain appreciable values of the transmission confined to a small range of frequencies. Our simulation shows that the structure of cylinders made from a ferromagnetic material can have left-handed properties under the action of a magnetic field *H* with low losses, while the parameters can be, furthermore, controlled by the applied magnetic field and the distance between cylinders.

This work has been supported by the Spanish DGICyT .

**Figure Captions**

Fig.1. (a) The magnetic permeability of a ferromagnetic material with $\omega_m/2p = 6$GHz, $\alpha=0.001$ and $\omega_0/2p = 18$GHz. (b) The effective dielectric permittivity of a Ni cylinder-structure with a diameter d=0.1cm and a distance between the cylinders a=0.5cm.

Fig.2. The transmission, reflection and absorption of the ferromagnetic cylinder structure with d=0.1cm and a=0.5cm for the $\omega_m/2p = 23$GHz, $\alpha=0.001$ and $\omega_0/2p = 18$GHz for (a) the normal incident wave, (b) for 45° angle of incidence and (c) for a random cylinders structure for a normal incident wave.

Fig.3. The transmission, reflection and absorption of the ferromagnetic cylinder structure with d=0.1cm and a=0.5cm for the $\omega_m/2p = 23$GHz, and $\omega_0/2p = 18$GHz for $\alpha=0.005$ and (a) three rows of the cylinders, (b) for 10 rows of cylinders.

Fig. 4a. The focusing properties of the material (3 rows of cylinders) for a point source located at 1 cm. Notice the focus at 0.65 cm corresponding to an effective real part of the refraction index of -1.1 with a very small imaginary part given the large transmitivity. The three rows of cylinders are the very white (zero field) of the figure. The frequency is 19GHz and characteristic magnetization 23GHz. (where is figure 4b)?

(a)

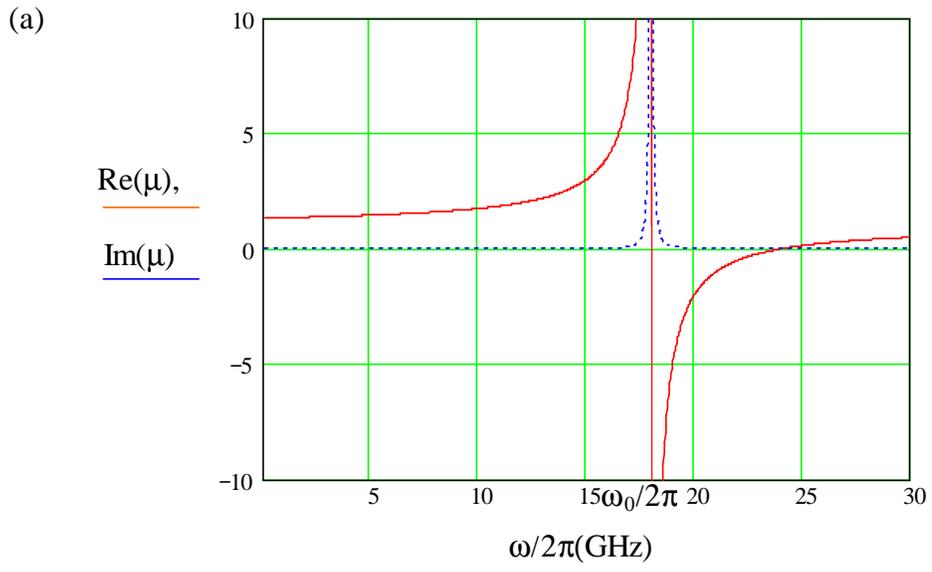

(b)

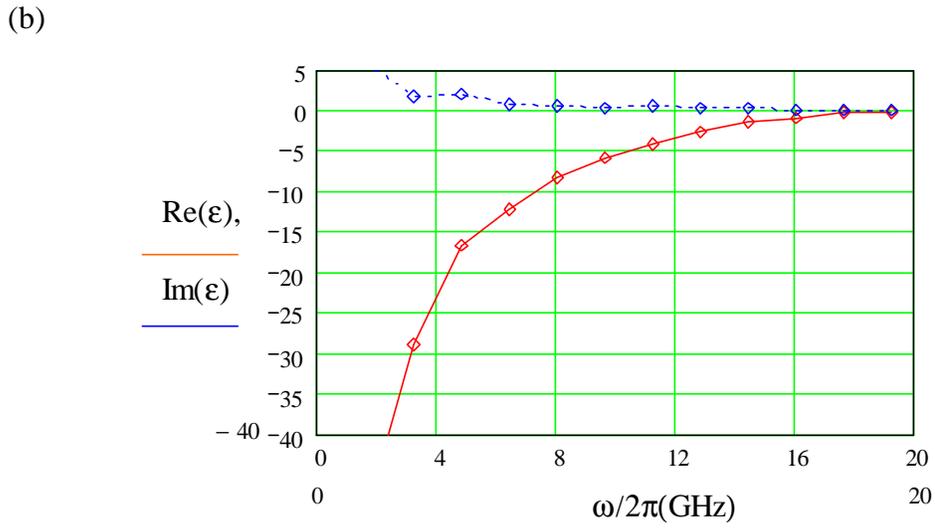

Fig.1

(a)

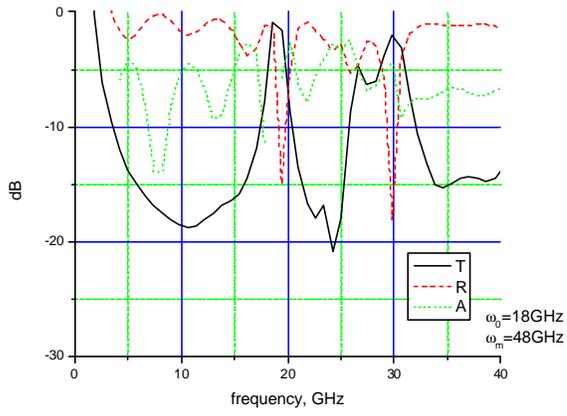

(b)

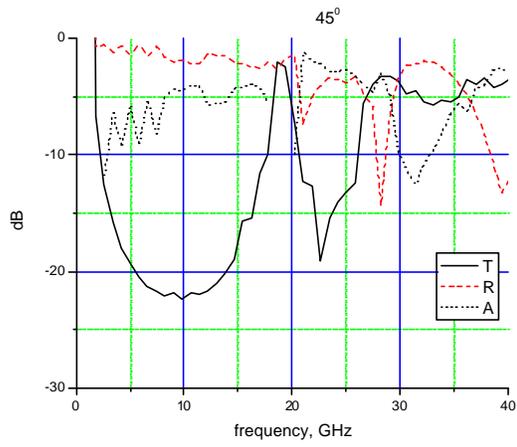

(c)

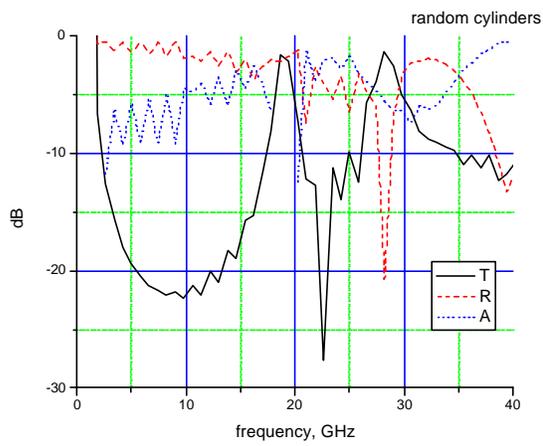

Fig.2

(a)

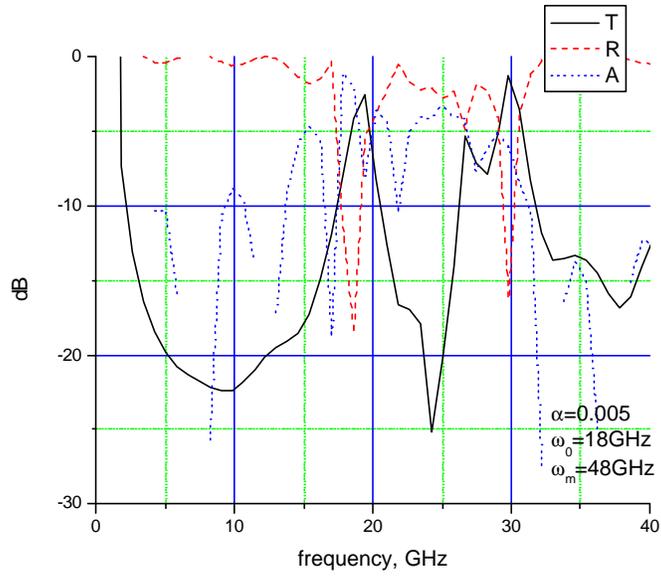

(b)

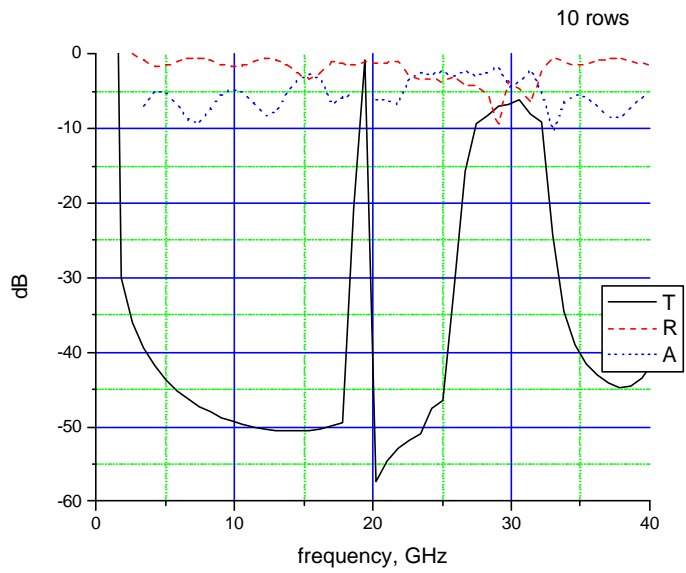

Fig.3

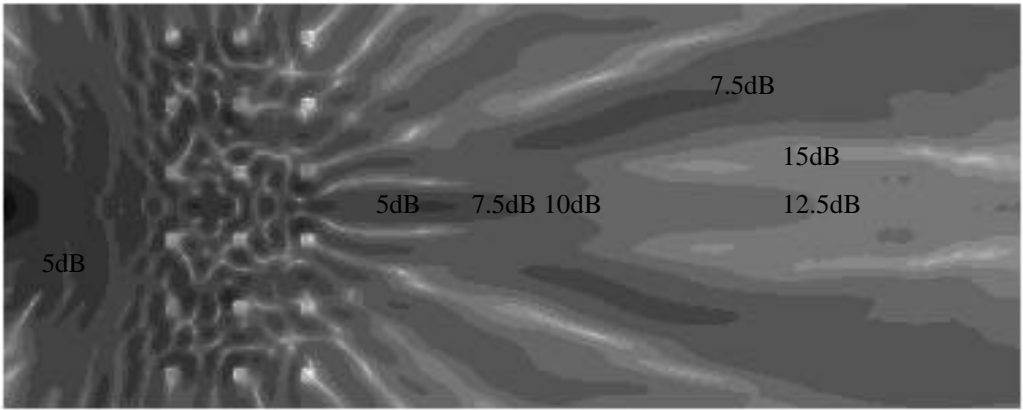

Fig.4